\documentclass[a4paper,12pt]{article}
\usepackage{amssymb}
\usepackage{tipa}
\usepackage{mathrsfs}
\usepackage{bbm}
\usepackage{epsf,amsmath}
\usepackage[dvips,usenames]{color}
\usepackage{graphicx}
\usepackage{color}
\usepackage{colortbl}

\newlength{\dinwidth}
\newlength{\dinmargin}
\begin{document}
\title{\bf The Effects of a Family Non-universal $Z^\prime$ Boson in the  $\bar{B}_s\to \pi K$, $\pi K^{\ast}$, $\rho K$ Decays and $B_d-\bar{B_d}$ Mixing}
\author{Qin Chang$^{a,b}$\footnote{Corresponding author. changqin@htu.cn}, Ya-Dong Yang$^{b,c}$\\
{ $^a$\small Department of Physics, Henan Normal University,
Xinxiang, Henan 453007, P.~R. China}\\
{$^{b}$\small Institute of Particle Physics, Huazhong Normal
University, Wuhan,
Hubei  430079, P. R. China}\\
{ $^c$\small Key Laboratory of Quark \& Lepton Physics, Ministry of
Education, P.R. China}}
\date{}
\maketitle
\bigskip\bigskip
\maketitle \vspace{-1.5cm}

\begin{abstract}
We revisit $\bar{B}_s\to \pi K$, $\pi K^{\ast}$ and $\rho K$ decays within QCDF formalism and examine the possible effects of a family non-universal $Z^\prime$ boson in these decays. In our evaluations, the strong constraints from $B_d-\bar{B_d}$ mixing on the strength of the left-handed flavor-changing $d-b-Z^{\prime}$ coupling $B_{db}^{L}$~($\sim0.2\times10^{-3}$) is also included. Numerically, we find that a new weak phase $\phi^L_d\sim-50^{\circ}$ involved in $B_{db}^{L}$, the negative combination of the $u-u-Z^{\prime}$ and $d-d-Z^{\prime}$ couplings  $P_{ud}^{R}$ and/or $D_{ud}^{L}$ with larger absolute value are crucial to improve the agreement of ${\cal B}(\bar{B}_s\to \pi^- K^+)$ between the SM prediction and the experimental measurement. Generally, ${\cal B}(\bar{B}_s\to \pi^- K^+)$ could be reduced by about $18\%$ at most by $Z^{\prime}$ contributions. Moreover, combining with the recent updated measurement on $\beta_s$ by CDF and D0 collaborations, we also find that $Z^{\prime}$ effects in $\bar{B}_s\to \pi^- K^+$ decays are very important for examining whether the CKM-like hierarchy persists in a family non-universal $Z^{\prime}$ coupling matrix.
\end{abstract}

\newpage
\section{Introduction}
The fruitful running of BABAR, Belle, CDF and D0 in the past years provides a very fertile testing ground for the Standard Model~(SM) picture of flavor physics and CP violations. Although most of the measurements are in perfect agreement with the SM predictions, there still exist some unexplained mismatches, such as the so-called ``$\pi K$, $\pi\pi$ puzzles'', anomalous $\bar{B}_s-B_s$ mixing phase, large transversal polarization in $B\to\phi K^{\ast}$, mismatching forward-backward asymmetry in $B\to K^{\ast}\mu^+\mu^-$ and so on. Recently, some interesting measurements of  $B_s$ decays have been reported by CDF collaboration at Fermilab Tevatron, which have attracted many attentions. Especially, CDF collaboration has made the first measurement of $\bar{B}_s\to \pi^- K^+$ decay~\cite{CDF1,CDF2}
\begin{eqnarray}\label{CDFExp}
\left\{\begin{array}{l}
 {\cal B}(\bar{B}_s\to \pi^- K^+)=(5.0\pm0.7\pm0.8)\times10^{-6}\,,\\
 A^{dir}_{cp}(\bar{B}_s\to \pi^- K^+)=(39\pm15\pm8)\%\,.
\end{array}\right.~{\rm CDF~Collaboration}
\end{eqnarray}

The recent theoretical
predictions for these two quantities based on the QCD factorization
approach~(QCDF)\cite{Beneke1}, the perturbative QCD
approach~(pQCD)\cite{KLS} and the soft-collinear effective
theory~(SCET)~\cite{scet} read
%%%%%%%%%%%%%%%%%%%%%%%%%%%%%%%%%%%%%%%%%%%%%%%%%
\begin{eqnarray}
\label{QCDFValue}
&&\left\{\begin{array}{l}
{\cal B}(\bar{B}_s\to \pi^- K^+)_{QCDF}=8.3\times10^{-6}\,, \\
A^{dir}_{cp}(\bar{B}_s\to \pi^- K^+)_{QCDF}=10.9\%~;
\end{array}\right.~{\rm QCDF~Scenario~S4}~\cite{Beneke3}\\
\label{pQCDValue} &&\left\{\begin{array}{l}
{\cal B}(\bar{B}_s\to \pi^- K^+)_{pQCD}=(7.6^{+3.2}_{-2.3}\pm0.7\pm0.5)\times10^{-6}\,, \\
A^{dir}_{cp}(\bar{B}_s\to \pi^- K^+)_{pQCD}=(24.1^{+3.9+3.3+2.3}_{-3.6-3.0-1.2})\%\,;
\end{array}\right.~{\rm pQCD}~\cite{PikPQCD}\\
\label{SCETValue} &&\left\{\begin{array}{l}
{\cal B}(\bar{B}_s\to \pi^- K^+)_{SCET}=(4.9\pm1.2\pm1.3\pm0.3)\times10^{-6}\,, \\
A^{dir}_{cp}(\bar{B}_s\to \pi^- K^+)_{SCET}=(20\pm17\pm19\pm5)\%.
\end{array}\right. {\rm SCET}~\cite{PikSCET}
\end{eqnarray}
%%%%%%%%%%%%%%%%%%%%%%%%%%%%%%%%%%%%%%%%%%%%%%%%%
One may note that QCDF~(Scenario~S4) result for ${\cal B}(\bar{B}_s\to \pi^- K^+)$ is significantly larger than CDF measurement Eq.~(\ref{CDFExp}). The pQCD prediction agree with the data in Eq.~(\ref{CDFExp}) within its large theoretical uncertainty, even though its central value also much larger than the measurement Eq.~(\ref{CDFExp}). In SCET, the contributions involving internal charm quark loops are claimed to be non-perturbative and only can be determined by fitting to the
data. A recent fit analysis of charmless B decays using SCET gives the result Eq.~(\ref{SCETValue}), which agrees with CDF data.
All of the theoretical predictions for $A^{dir}_{cp}(\bar{B}_s\to \pi^- K^+)$ agree with CDF data with their large theoretical uncertainties.

For this possible branching ratio mismatch, many  efforts have been done within both SM and possible New Physics~(NP) scenarios, for example Refs.~\cite{SU3ana,GHZhu,HYCheng,RMWang}. A recent detailed analysis in Ref.~\cite{SU3ana} indicates that $SU(3)$ and factorization only remain approximately valid if the branching ratio for $\bar{B}_s\to \pi^- K^+$ exceeds its current value by at least $42\%$, or if a parameter
$\xi$ describing ratios of form factors and decay constants is shifted from its nominal value by more than
twice its estimated error. Such analysis also implies that, if ${\cal B}_{Exp}(\bar{B}_s\to \pi^- K^+)$ and $\xi$ persist in the future, large ${\cal B}(\bar{B}_s\to \pi^- K^+)$ mismatch is possibly induced by NP effects.

A new family non-universal $Z^\prime$ boson could be naturally derived in certain
string constructions~\cite{string}, $E_6$ models~\cite{E6} and so
on. Searching for such an extra $Z^{\prime}$ boson is an important
mission in the experimental programs of Tevatron~\cite{Tevatron} and
LHC~\cite{LHC}. Within a model-independent formalism for a family non-universal $Z^{\prime}$ model presented by Ref.~\cite{Langacker}, $b\to s$ transition involving $b-s-Z^{\prime}$ coupling has been widely studied. It is interesting that such $Z^\prime$ boson behavior is helpful to resolve many puzzles in $b\to s$ transition, such as ``$\pi K$ puzzle''~\cite{Barger1,Chang1} and anomalous $\bar{B}_s-B_s$ mixing phase~\cite{Barger2,Chang2}. So, it is worth evaluating the effects of a non-universal $Z^\prime$ boson on $b\to d$ transition, especially the measured $\bar{B}_s\to \pi^- K^+$. Moreover, with the constraints from $\bar{B}_q-B_q$ mixing, $B\to\pi K$, $\pi K^{\ast}$, $\rho K$ and $X_s\mu^{+}\mu^{-}$ decays, a CKM-like hierarchy persists in $Z^\prime$ coupling matrix (at least between $b-s-Z^{\prime}$ and $b-d-Z^{\prime}$ couplings)~\cite{Chang2}. Naturally, it is important to examine whether such hierarchy is persisted after the constraints from $\bar{B}_s\to \pi^- K^+$ are included.

In addition, the correlated decays $\bar{B}_s\to\pi K^{\ast}$ and $\rho K$, which also involves $b\to d\bar{q} q$~($q=u,d$) transition at quark level, will be measured at the running LHC. Such decays may perform an important role in testing SM and searching for NP signal. Especially, due to large theoretical uncertainty cancellation, the ratio ${\cal B}(\bar{B}_s\to \rho^- K^+)/{\cal B}(\bar{B}_s\to \pi^- K^+)$ is much more suitable for testing SM and probing NP~\cite{GHZhu}. Therefore, we also evaluate $\bar{B}_s\to\pi K^{\ast}$ and $\rho K$ decays within both SM and a family non-universal $Z^{\prime}$ model.

Our paper is organized as follows. In Section 2, we revisit $\bar{B}_s\to \pi K$, $\pi K^{\ast}$ and $\rho K$ decays in the SM within the QCDF formalism; we mainly evaluate the  effects of annihilation contributions which involve endpoint divergency. In Section 3, we perform a fitting on $b-d-Z^{\prime}$ couplings with the constraints from $\bar{B}_d-B_d$ mixing and $\bar{B}_s\to \pi^- K^+$ decay; the NP effects on the other decays are also evaluated. Section 4 is our conclusions.  Theoretical input parameters are summarized in the Appendix.

\section{Revisiting $\bar{B}_s\to \pi K$, $\pi K^{\ast}$ and $\rho K$  Decays within QCDF Formalism}

In the SM, the effective weak Hamiltonian responsible for $b\to d$
transitions is given as~\cite{Buchalla:1996vs}
%%%%%%%%%%%%%%%%%%%%%%%%%%%%%%%%%%%%%%%%%%%%%%%%%
\begin{eqnarray}\label{eq:eff}
 {\cal H}_{\rm eff} &=& \frac{G_F}{\sqrt{2}} \biggl[V_{ub}
 V_{ud}^* \left(C_1 O_1^u + C_2 O_2^u \right) + V_{cb} V_{cd}^* \left(C_1
 O_1^c + C_2 O_2^c \right) - V_{tb} V_{td}^*\, \big(\sum_{i = 3}^{10}
 C_i O_i \big. \biggl. \nonumber\\
 && \biggl. \big. + C_{7\gamma} O_{7\gamma} + C_{8g} O_{8g}\big)\biggl] +
 {\rm h.c.},
\end{eqnarray}
%%%%%%%%%%%%%%%%%%%%%%%%%%%%%%%%%%%%%%%%%%%%%%%%%
where $V_{qb} V_{qd}^*$~($q=u, c$ and $t$) are products of the
Cabibbo-Kobayashi-Maskawa~(CKM) matrix elements~\cite{ckm}, $C_{i}$
the Wilson coefficients, and $O_i$ the relevant four-quark operators.

In recent years, QCDF has been employed extensively to study the B
meson non-leptonic decays. For example, all of the decay modes
considered here have been studied comprehensively within the SM in
Refs.~\cite{Beneke3,HYCheng}. The relevant decay
amplitudes for $\bar{B}_s\to \pi K$, $\pi K^{\ast}$ and $\rho K$ decays
within the QCDF formalism are
%%%%%%%%%%%%%%%%%%%%%%%%%%%%%%%%%%%%%%%%%%%%%%%%%
\begin{eqnarray}\label{amplitud}
 {\cal A}_{\bar{B}_s\to \pi^- K^+}&=& A_{K\pi}
 \left[
 \delta_{pu}\alpha_1+\alpha_4^p+\alpha_{4,\rm EW}^p+\beta_3^p-\frac{1}{2}\beta_{3,\rm EW}^p
 \right]\,,\nonumber\\
 \sqrt{2}{\cal A}_{\bar{B}_s\to \pi^0 K^0}&=& A_{K\pi}
 \left[ \delta_{pu}\alpha_2-\alpha_4^p+\frac{3}{2}\alpha_{3,\rm EW}^p+\frac{1}{2}\alpha_{4,\rm EW}^p-\beta_3^p+\frac{1}{2}\beta_{3,\rm EW}^p
 \right]\,.
\end{eqnarray}
%%%%%%%%%%%%%%%%%%%%%%%%%%%%%%%%%%%%%%%%%%%%%%%%%
where the explicit expressions for the coefficients
$\alpha_i^p\equiv\alpha_i^p(M_1M_2)$ and
$\beta_i^p\equiv\beta_i^p(M_1M_2)$ can also be found in
Ref.~\cite{Beneke3}. The expressions for the $\bar{B}_s\to\pi K^{\ast}$ and $\rho K$ amplitudes are obtained by setting $(\pi K)\to(\pi K^{\ast})$ and $(\pi K)\to(\rho K)$,
 respectively.

It is noted that the QCDF framework contains estimates of some power-suppressed but
numerically important contributions, such as the annihilation
corrections. Unfortunately, in a collinear factorization approach, the calculation of annihilation amplitude always suffers from end-point divergence. In the pQCD approach, such divergence is regulated by the parton transverse momentum $k_T$ at the expense of modeling additional $k_T$ dependence of meson distribution functions~\cite{KLS},
and a large strong phase is found. In Refs.~\cite{PikSCET,Feldmann}, annihilation diagram is studied with SCET and parameterized by a complex amplitude. At present, the dynamical origin of these corrections still remains a theoretical challenge. Within the QCDF framework, in a model-dependent way, there are two schemes to regulate the endpoint divergence until now.
Scheme~I: In Ref.~\cite{Beneke3}, to probe their possible effects
conservatively, the endpoint divergent integrals are treated as signs of infrared sensitive contribution
 and  phenomenologically parameterized by
%%%%%%%%%%%%%%%%%%%%%%%%%%%%%%%%%%%%%%%%%%%%%%%%%
\begin{equation}\label{treat-for-anni}
\int_0^1 \frac{\!dx}{x}\, \to X_{A,H} =(1+\rho_{A,H} e^{i\phi_{A,H}}) \ln
\frac{m_B}{\Lambda_h}, \qquad \int_0^1dy \frac{\textmd{ln}y}{y}\,
\to -\frac{1}{2}(X_{A,H})^2
 \end{equation}
%%%%%%%%%%%%%%%%%%%%%%%%%%%%%%%%%%%%%%%%%%%%%%%%%
with $\rho_{A,H} \leq 1$ and $\phi_{A,H}$ unrestricted. The different
scenarios corresponding to different choices of $\rho_{A,H}$ and
$\phi_{A,H}$ have been thoroughly discussed in Ref.~\cite{Beneke3}.
Scheme~II: As an alternative scheme to the first one, one could quote an infrared finite gluon propagator to regulate the endpoint divergent integrals, which have been thoroughly studied in Ref.~\cite{Chang3}.

%%%%%%%%%%%%%%%%%%%%%%%%%%%%%%%%%%%%%%%%
\begin{figure}[ht]
\begin{center}
\epsfxsize=15cm \centerline{\epsffile{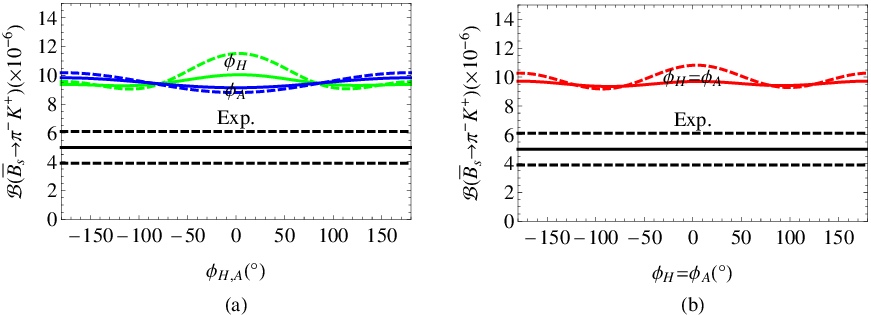}}
\centerline{\parbox{16cm}{\caption{\label{FigBrphi}\small The dependence of ${\cal B}(\bar{B}_s\to \pi^- K^+)$ on $\phi_{A,H}$ with $\rho_{A,H}=1$~(colored real line), $2$(colored dashed line). The black horizontal lines are the experimental data with the solid line being the central value and the dashed ones the error-bars (1$\sigma$). }}}
\end{center}
\end{figure}
%%%%%%%%%%%%%%%%%%%%%%%%%%%%%%%%%%%%%%%%%%%

In scheme~I, with the central value of the input parameters listed in the Appendix, Fig.~\ref{FigBrphi} shows the dependence of ${\cal B}(\bar{B}_s\to \pi^- K^+)$ on $\phi_{A(H)}$ with $\rho_{A(H)}=1$, $2$ and $\rho_{H(A)}=0$.
As Fig.~\ref{FigBrphi}(a) shows, assuming $\phi_{A}\neq\phi_{H}$, the minimal ${\cal B}(\bar{B}_s\to \pi^- K^+)$ appears at $\phi_{H}\simeq0^{\circ}$ and $\phi_{A}\simeq-110^{\circ}$; Assuming $\phi_{A}=\phi_{H}$, it appears at $\phi_{H}=\phi_{A}\simeq-100^{\circ}$, as shown in Fig.~\ref{FigBrphi}~(b). With the above $\phi_{H,A}$ values, $\rho_{A,H}=1$ and the central values of the other inputs, corresponding above two assumptions for $\phi_{A,H}$, we get ${\cal B}(\bar{B}_s\to \pi^- K^+)=9.0\times10^{-6}$, $9.3\times10^{-6}$ respectively, which are also significantly larger than CDF data $(5.0\pm1.1)\times10^{-6}$.
Moreover, corresponding to above two assumptions, we find the direct CP asymmetry for $\bar{B}_s\to \pi^- K^+$ is $3\%$, $6\%$ respectively, which are also lower than CDF measurement $(39\pm17)\%$ at $2\sigma$  level. So, following a reasonable choice presented by Cheng and Chua~\cite{HYCheng}, in our following calculations, we take $\rho_{A,H}(PP,PV,VP)=1$ and $\phi_{A,H}=-55^{\circ}(PP)$, $-30^{\circ}(PV)$, $-65^{\circ}(VP)$. Such choice of the values of parameters $\rho_{A,H}$ and $\phi_{A,H}$ are similar to the so-called favored scenario ``scenario S4'' for $B_{u,d}$ decays presented by Beneke and Neubert~\cite{Beneke3}, except that some additional moderate values of input parameters (such as $m_s=80\,MeV$ and $\alpha_2^{\pi}=0.3$) are also used in ``scenario S4''~\cite{Beneke3}.  For estimating theoretical uncertainties, we shall assign an error of $\pm 0.1$ to $\rho_{A,H}$ and $\pm 5^{\circ}$ to $\phi_{A,H}$. With such $\rho_{A,H}$, $\phi_{A,H}$ values and the other inputs listed in the Appendix,  we present the QCDF predictions for $\bar{B}_s\to \pi K$, $\pi K^{\ast}$ and $\rho K$ decays in the third column of Table~\ref{br}, \ref{Acp} and \ref{mix}.

%%%%%%%%%%%%%%%%%%%%%%%%%%%%%%%%%%%%%%%%
\begin{table}[t]
 \begin{center}
 \caption{The numerical results for ${\cal B}(\bar{B}_s\to \pi K,\pi K^{\ast},\rho K)(\times10^{-6})$.}
 \label{br}
 \vspace{0.5cm}
 \small
 \doublerulesep 0.1pt \tabcolsep 0.05in
 \begin{tabular}{lccccccccccc} \hline \hline
                                         &\multicolumn{1}{c}{Exp.}&\multicolumn{2}{c}{SM}            &\multicolumn{2}{c}{SM$+Z^{\prime}$}\\
                                         &            & Scheme I             & Scheme II             & S1                    & S2      \\ \hline
 ${\cal B}(\bar{B}_s\to \pi^- K^+)$      &$5.0\pm1.1$ &$9.5^{+3.8}_{-2.6}$   &$10.2^{+3.7}_{-2.5}$   &$8.6^{+3.8}_{-2.8}$    &$8.9^{+3.9}_{-2.6}$        \\
 ${\cal B}(\bar{B}_s\to \pi^0 K^0)$      &---         &$0.53^{+0.35}_{-0.12}$&$0.39^{+0.23}_{-0.08}$ &$0.50^{+0.40}_{-0.17}$ &$0.52^{+0.41}_{-0.20}$\\
 \hline
 ${\cal B}(\bar{B}_s\to \pi^- K^{\ast+})$&---         &$13.3^{+4.5}_{-3.4}$  &$14.1^{+4.0}_{-3.4}$   &$14.0^{+4.6}_{-3.9}$   &$13.7^{+5.2}_{-3.6}$    \\
 ${\cal B}(\bar{B}_s\to \pi^0 K^{\ast0})$ &---         &$0.38^{+0.23}_{-0.15}$&$0.28^{+0.16}_{-0.06}$&$0.21^{+0.27}_{-0.10}$ &$0.26^{+0.16}_{-0.15}$     \\
 \hline
 ${\cal B}(\bar{B}_s\to \rho^- K^+)$     &---         &$22.5^{+9.2}_{-6.5}$ &$24.5^{+9.1}_{-6.4}$    &$22.2^{+10.2}_{-6.6}$   &$22.2^{+9.2}_{-6.4}$       \\
 ${\cal B}(\bar{B}_s\to \rho^0 K^0)$     &---         &$0.79^{+0.42}_{-0.35}$&$0.27^{+0.09}_{-0.06}$ &$0.59^{+0.66}_{-0.37}$ &$0.64^{+0.58}_{-0.37}$        \\
 \hline
 \hline \hline
 \end{tabular}
 \end{center}
 \end{table}
%%%%%%%%%%%%%%%%%%%%%%%%%%%%%%%%%%%%%%%%%%%%

%%%%%%%%%%%%%%%%%%%%%%%%%%%%%%%%%%%%%%%%
\begin{table}[t]
 \begin{center}
 \caption{The numerical results for $A^{dir}_{cp}(\bar{B}_s\to \pi K,\pi K^{\ast},\rho K)(\times10^{-2})$.}
 \label{Acp}
 \vspace{0.5cm}
 \small
 \doublerulesep 0.1pt \tabcolsep 0.05in
 \begin{tabular}{lccccccccccc} \hline \hline
                                             &\multicolumn{1}{c}{Exp.}&\multicolumn{2}{c}{SM}      &\multicolumn{2}{c}{SM$+Z^{\prime}$}\\
                                             &                        & Scheme I              & Scheme II             & S1                   & S2      \\ \hline
 $A^{dir}_{cp}(\bar{B}_s\to \pi^- K^+)$      &$39\pm17$               &$13.6^{+6.4}_{-4.6}$   &$22.4^{+7.6}_{-5.7}$   &$14.2^{+6.8}_{-5.7}$  &$14.0^{+7.6}_{-5.2}$        \\
 $A^{dir}_{cp}(\bar{B}_s\to \pi^0 K^0)$      &---                  &$19.7^{+10.6}_{-15.5}$ &$10.4^{+7.9}_{-7.8}$ &$25.7^{+15.5}_{-16.0}$ &$23.8^{+16.3}_{-14.3}$\\
 \hline
 $A^{dir}_{cp}(\bar{B}_s\to \pi^- K^{\ast+})$&---                     &$-17.1^{+5.7}_{-4.8}$  &$-23.9^{+7.7}_{-6.5}$  &$-16.5^{+5.7}_{-4.3}$ &$-16.7^{+5.7}_{-5.1}$    \\
 $A^{dir}_{cp}(\bar{B}_s\to \pi^0 K^{\ast0})$&---                     &$-15.0^{+10.6}_{-26.1}$&$-16.2^{+7.1}_{-13.8}$ &$-50^{+42}_{-34}$     &$-34.9^{+30}_{-28}$     \\
 \hline
 $A^{dir}_{cp}(\bar{B}_s\to \rho^- K^+)$     &---                     &$9.6^{+2.7}_{-5.4}$    &$18.3^{+4.6}_{-4.8}$   &$9.6^{+1.8}_{-5.4}$   &$9.6^{+1.7}_{-5.2}$       \\
 $A^{dir}_{cp}(\bar{B}_s\to \rho^0 K^0)$     &---                 &$3.0^{+18.8}_{-8.5}$   &$-19.9^{+16.3}_{-10.9}$&$23.7^{+44}_{-23}$    &$16.0^{+38.2}_{-18.8}$        \\
 \hline
 \hline \hline
 \end{tabular}
 \end{center}
 \end{table}
%%%%%%%%%%%%%%%%%%%%%%%%%%%%%%%%%%%%%%%%%%%%

%%%%%%%%%%%%%%%%%%%%%%%%%%%%%%%%%%%%%%%%
\begin{table}[t]
 \begin{center}
 \caption{The numerical results for $A^{mix}_{CP}(\bar{B}_s\to \pi^0 K_s)$ and $A^{mix}_{CP}(\bar{B}_s\to \rho^0 K^0)$$(\times10^{-2})$.}
 \label{mix}
 \vspace{0.5cm}
 \small
 \doublerulesep 0.1pt \tabcolsep 0.05in
 \begin{tabular}{lccccccccccc} \hline \hline
                                      &\multicolumn{1}{c}{Exp.}&\multicolumn{2}{c}{SM}     &\multicolumn{2}{c}{SM$+Z^{\prime}$}\\
                                      &                        & Scheme I         & Scheme II          & S1          & S2             \\ \hline
 $A^{mix}_{CP}(\bar{B}_s\to \pi^0 K_s)$     &---                     &$-83^{+19}_{-13}$ &$-96^{+2.9}_{-1.8}$ &$-89^{+23}_{-8}$   &$-88^{+20}_{-93}$        \\
 $A^{mix}_{CP}(\bar{B}_s\to \rho^0 K^0)$    &---                     &$-22^{+35}_{-28}$ &$-78^{+12}_{-7}$    &$-56^{+59}_{-31}$  &$-46^{+48}_{-41}$\\
 \hline
 \hline \hline
 \end{tabular}
 \end{center}
 \end{table}
%%%%%%%%%%%%%%%%%%%%%%%%%%%%%%%%%%%%%%%%%%%%

In scheme~II, we adopt the infrared finite gluon propagator derived by Cornwall~\cite{Cornwall}, which involves a new parameter gluon mass scale $m_g$, to regulate the endpoint divergency. In Ref.~\cite{Chang3}, we present our suggestion $m_g\sim500~{\rm MeV}$, which is a reasonable choice so that most of the observables for $B\to\pi K$,~$\pi K^{\ast}$ and $\rho K$ decays are in good agreement with the experimental data. Furthermore, compared to the available data for $B^0\to K^+K^-$,$D_s^{(\ast)}K$, $B_s\to\pi\pi$ and so on, the most recent evaluation~\cite{Zanetti} also presents that the gluon mass scale is close to $500~{\rm MeV}$. So, in this paper, we shall take $m_g=500~{\rm MeV}$. With such a value, the predictions for observables of $\bar{B}_s\to \pi K$, $\pi K^{\ast}$ and $\rho K$ decays are listed in the fourth column in Table~\ref{br}, \ref{Acp} and \ref{mix}. We find most of the predictions in scheme~I and scheme~II are consistent with each other except for some observables in $\bar{B}_s\to \rho^0 K^0$, $\pi^0 K^{\ast0}$ decays. Because the cancellations among $\alpha_2$ and $-\alpha_4^p+\frac{3}{2}\alpha_{3,\rm EW}^p+\frac{1}{2}\alpha_{4,\rm EW}^p$ in Eq.~(\ref{amplitud}) for $\bar{B}_s\to \rho^0 K^0$, $\pi^0 K^{\ast0}$ decays, the effect of annihilation contributions is significant. Furthermore, in $B\to PV$ decay, the annihilation contributions in scheme~II provide a larger imaginary part than the one in scheme~I~\cite{Chang3}.
So, the two schemes present some different predictions for $\bar{B}_s\to \rho^0 K^0$, $\pi^0 K^{\ast0}$ decays, which will be judged by the upcoming  LHC-b and proposed super-B experiments.

Within both schemes~I and II, we find that ${\cal B}_{SM}(\bar{B}_s\to \pi^- K^+)$ is significantly larger than the measurement $(5.0\pm1.1)\times10^{-6}$. Due to the dominance of the tree contribution $\alpha_1$ in the amplitude for $\bar{B}_s\to \pi^- K^+$ decay, the contributions from annihilations are tiny in comparison to  $\alpha_1$. So, the large ${\cal B}(\bar{B}_s\to \pi^- K^+)$
problem could not be resolved by
tuning the  annihilation contributions.

A recent detailed analysis~\cite{SU3ana} in a flavor symmetry framework indicate that the mismatch ${\cal B}(\bar{B}_s\to \pi^- K^+)$ problem persists within the SM if experimental result of ${\cal B}(\bar{B}_s\to \pi^- K^+)$ would not exceeds its current value $(5.0\pm1.1)\times10^{-6}$~\cite{CDF1,CDF2} by at least $50\%$ or if the SU(3)-breaking factor defined by
\begin{equation}
 \xi\equiv\frac{f_K F_{B^0\pi}(m_K^2)}{f_{\pi} F_{B_sK}(m_{\pi}^2)}\frac{m_{B^0}^2-m_{\pi}^2}{m_{B_s}^2-m_{K}^2}\,
\end{equation}
is not shifted from the value $0.97^{+0.09}_{-0.11}$~\cite{SU3ana,su3ana2}~(corresponding to almost exact $SU(3)$, see Refs.~\cite{SU3ana,su3ana2} for detail) by more than twice its estimated error. With the central value $F_0^{B_sK}=0.24$ obtained by lattice and pQCD calculations, recent QCDF analysis~\cite{HYCheng} gets a good result ${\cal B}(\bar{B}_s\to \pi^- K^+)\approx5.3\times10^{-6}$. However, as mentioned in Ref.~\cite{HYCheng}, their result $\xi=1.24$ is larger than $0.97$ at $3\sigma$. Furthermore, $F_0^{B_sK}$ is predicted by lattice QCD computations at small recoil, while unfortunately charmless B decays happen at large recoil.
So, if the future refined measurement and theoretical study confirm the large ${\cal B}(\bar{B}_s\to \pi^- K^+)$ problem, it will be a significant signal of NP. In our following study, we will examine the effects of a family non-universal $Z^{\prime}$ model for possible solution.

\section{The effects of  family non-universal $Z^{\prime}$ model}
\subsection{Brief review of the family non-universal $Z^{\prime}$ model}
A possible heavy $Z^{\prime}$ boson is predicted in many extensions
of the SM, such as grand unified theories, superstring theories, and
theories with large extra dimensions. The simplest way to extend the
SM gauge structure is to include a new $U(1)$ gauge group. A family
non-universal $Z^{\prime}$ boson can lead to FCNC processes even at
tree level due to the non-diagonal chiral coupling matrix. The
formalism of the model has been detailed in Ref.~\cite{Langacker}.
The relevant studies in the context of B physics have also been
extensively performed in Refs.~\cite{Barger1,Barger2,BZprime,Chang4}.

With the assumption of flavor-diagonal right-handed couplings , the
$Z^{\prime}$ part of the effective Hamiltonian for $b\to
d\bar{q}q$ $(q=u,d)$ transitions can be written as~\cite{Chang1,Barger2}
%%%%%%%%%%%%%%%%%%%%%%%%%%%%%%%%%%%%%%%%%%%%%%%%%
\begin{equation}
 {\cal H}_{eff}^{\rm
 Z^{\prime}}=-\frac{G_F}{\sqrt{2}}V_{td}^{\ast}V_{tb}\sum_{q}
 (\Delta C_3 O_3^q +\Delta C_5 O_5^q+\Delta C_7 O_7^q+\Delta C_9
  O_9^q)+h.c.\,,
\end{equation}
%%%%%%%%%%%%%%%%%%%%%%%%%%%%%%%%%%%%%%%%%%%%%%%%%
where $O_i^q(i=3,5,7,9)$ are the effective operators in the SM, and
$\Delta C_i$ the modifications to the corresponding SM Wilson
coefficients caused by $Z^{\prime}$ boson. In terms of the model parameters at
the $M_W$ scale,    $\Delta C_i$ are expressed as
%%%%%%%%%%%%%%%%%%%%%%%%%%%%%%%%%%%%%%%%%%%%%%%%%
\begin{eqnarray}
 \Delta C_{3,5}&=&-\frac{2}{3V_{td}^{\ast}V_{tb}}\,B_{db}^L\,(B_{uu}^{L,R}+2B_{dd}^{L,R})\,,\nonumber\\
 \Delta C_{9,7}&=&-\frac{4}{3V_{td}^{\ast}V_{tb}}\,B_{db}^L\,(B_{uu}^{L,R}-B_{dd}^{L,R})\,,
 \label{NPWilson}
\end{eqnarray}
%%%%%%%%%%%%%%%%%%%%%%%%%%%%%%%%%%%%%%%%%%%%%%%%%
where   $B_{q^{\prime}q}^{L,R}$ is the effective chiral $Z^{\prime}$
coupling matrix element.

Generally, the diagonal elements of the effective coupling matrix
$B_{qq}^{L,R}$ are real as a result of the hermiticity of the
effective Hamiltonian. However, the off-diagonal one $B_{db}^L$
may contain a new weak phase $\phi^L_d$. Then, conveniently we can
represent $\Delta C_i$ as
%%%%%%%%%%%%%%%%%%%%%%%%%%%%%%%%%%%%%%%%%%%%%%%%%
\begin{eqnarray}
 \Delta
 C_{3,5}&=&2\,\frac{|V_{td}^{\ast}V_{tb}|}{V_{td}^{\ast}V_{tb}}\,
 \zeta^{LL,LR}\,e^{i\phi^L_d}\,,\nonumber\\
 \Delta
 C_{9,7}&=&4\,\frac{|V_{td}^{\ast}V_{tb}|}{V_{td}^{\ast}V_{tb}}\,
 \xi^{LL,LR}\,e^{i\phi^L_d}\,,
\end{eqnarray}
%%%%%%%%%%%%%%%%%%%%%%%%%%%%%%%%%%%%%%%%%%%%%%%%%
where the real NP parameters $\zeta^{LL,LR}$, $\xi^{LL,LR}$ and
$\phi^L_d$ are defined, respectively, as
%%%%%%%%%%%%%%%%%%%%%%%%%%%%%%%%%%%%%%%%%%%%%%%%%
\begin{eqnarray}\label{comPara}
 \zeta^{LL,LR}&=&-\frac{1}{3}\,\big|\frac{B_{db}^L}{V_{td}^{\ast}V_{tb}}\big|\,P_{ud}^{L,R}\,,\quad P_{ud}^{L,R}= (B_{uu}^{L,R}+2B_{dd}^{L,R})\,;\nonumber\\
 \xi^{LL,LR}&=&-\frac{1}{3}\,
 \big|\frac{B_{db}^L}{V_{td}^{\ast}V_{tb}}\big|\,D_{ud}^{L,R}\,,\quad D_{ud}^{L,R}=(B_{uu}^{L,R}-B_{dd}^{L,R})\,;
 \nonumber\\
 \phi^L_d&=&{\rm Arg}[B_{db}^L]\,.
\end{eqnarray}
%%%%%%%%%%%%%%%%%%%%%%%%%%%%%%%%%%%%%%%%%%%%%%%%%

%%%%%%%%%%%%%%%%%%%%%%%%%%%%%%%%%%%%%%%%
\begin{table}[t]
 \begin{center}
 \caption{The inputs of the $Z^{\prime}$ couplings~\cite{Chang2}.}
 \label{inputsZp}
 \vspace{0.5cm}
 \small
 \doublerulesep 0.1pt \tabcolsep 0.05in
 \begin{tabular}{lccccccccccc} \hline \hline
 Solutions &$P_{ud}^{L}$  &$P_{ud}^{R}$ &$D_{ud}^{L}$ &$D_{ud}^{R}$\\\hline
 S1        &$2.1\pm3.3$   &$-0.3\pm1.3$ &$-0.5\pm0.4$  &$0.03\pm0.10$\\\hline
 S2        &$0.7\pm1.9$   &$-0.1\pm0.8$  &$-0.3\pm0.2$  &$0.01\pm0.05$\\
 \hline\hline
 \end{tabular}
 \end{center}
 \end{table}
%%%%%%%%%%%%%%%%%%%%%%%%%%%%%%%%%%%%%%%%%%%%

Firstly, we shall specify the values of the $Z^{\prime}$ input parameters in our calculations. In $b\to s\bar{q}q$ transition, the combinations of $b-s-Z^{\prime}$, $u-u-Z^{\prime}$ and $d-d-Z^{\prime}$ couplings~({\it i.~e.} $\zeta^{LL,LR}_s$ and $\xi^{LL,LR}_s$, which could be derived through replacing $B_{db}^L$ in  Eq.~(\ref{comPara}) by $B_{sb}^L$)  have been well bounded by the constraints from $B\to\pi K^{(\ast)}$ and $\rho K$ decays~\cite{Chang1}. After considering the constraints from $B_s-\bar{B}_s$ mixing, which performs constraint on $b-s-Z^{\prime}$
coupling solely, one may easily extract the results of $u-u-Z^{\prime}$ and $d-d-Z^{\prime}$ couplings $P_{ud}^{L,R}$ and $D_{ud}^{L,R}$~\cite{Chang2}, which are recapitulated in Table~\ref{inputsZp}\footnote{In ref.~\cite{Chang2}, the errors for the numerical results of the $Z^{\prime}$ couplings are obtained simply with a Gaussian distribution treatment of the outputs caused by the uncertainties of input parameters. In this paper, to evaluate the conservative effects of the $Z^{\prime}$ couplings, we would study the whole possible regions of the $Z^{\prime}$ couplings, which are listed in Table~\ref{inputsZp}.}. Such results are the same as the ones in $b\to d\bar{q}q$ transition and will be adopted in the following analysis.

As for  $b-d-Z^{\prime}$ coupling, in Ref.~\cite{Chang2}, we have found  the strength of $B_{db}^L$ is strongly suppressed by $B_d-\bar{B}_d$ mixing. So, in our following fitting, the constraints from $B_d-\bar{B}_d$ mixing are also considered. The basic theoretical framework for $B_q-\bar{B}_q$ mixing within a family non-universal $Z^{\prime}$ model have been given by Refs.~\cite{Barger2,Chang2} in detail. In this paper, we take the same framework and conventions as Ref.~\cite{Chang2}. Moreover, we also take the recent updated fitting results of two mixing parameters $C_{B_d}$ and $\phi_{B_d}$ by  UTfit collaboration \cite{UTfit},  which are defined as
\begin{equation}
\label{UTEq}
 C_{B_d}e^{2i\phi_{B_d}}\,\equiv\,\frac{\langle
 B_d|\mathcal{H}_{eff}^{full}|\bar{B}_d\rangle}
 {\langle B_d|\mathcal{H}_{eff}^{SM}|\bar{B}_d\rangle }\,
 =\,\frac{A_d^{SM}e^{i\phi_{d}^{SM}}+A_d^{NP}e^{i(2\phi_{d}^{NP}
 +\phi_{d}^{SM})}}{A_d^{SM}e^{i\phi_{d}^{SM}}}\,,
\end{equation}
and
\begin{equation}
C_{B_d}=0.95\pm0.14\,,\quad \phi_{B_d}=-3.1^{\circ}\pm1.7^{\circ}
\end{equation}
at $68\%$ probability\cite{UTfit}.

\subsection{Numerical analyses and discussions}
%%%%%%%%%%%%%%%%%%%%%%%%%%%%%%%%%%%%%%%%
\begin{figure}[ht]
\begin{center}
\epsfxsize=15cm \centerline{\epsffile{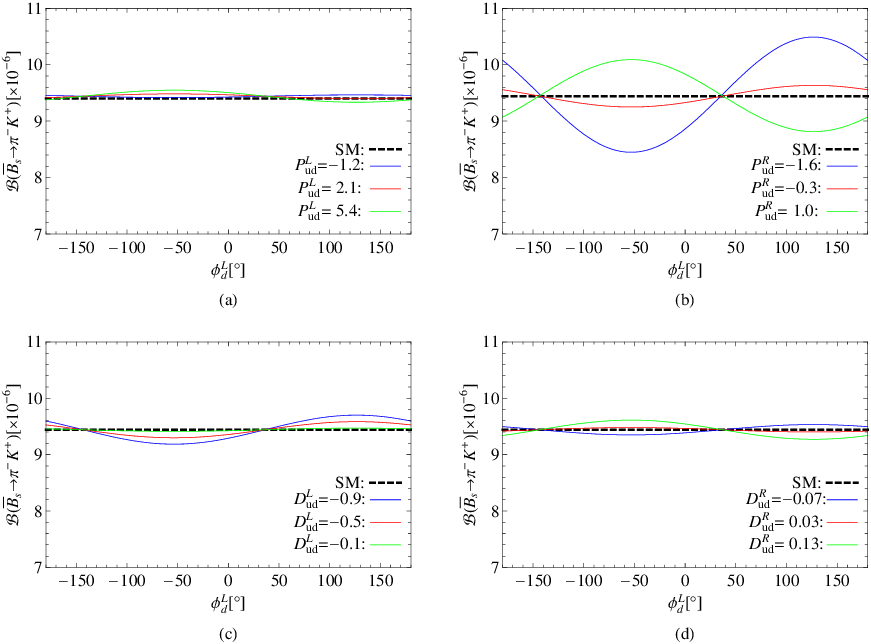}}
\centerline{\parbox{16cm}{\caption{\label{fig3}\small The dependence of ${\cal B}(\bar{B}_s\to \pi^- K^+)$ on $\phi^L_d$, $P_{ud}^{L,R}$ and $D_{ud}^{L,R}$ with $|B_{db}^L|=0.15(\times10^{-3})$ and the central values of theoretical input parameters. }}}
\end{center}
\end{figure}
%%%%%%%%%%%%%%%%%%%%%%%%%%%%%%%%%%%%%%%%%%%

%%%%%%%%%%%%%%%%%%%%%%%%%%%%%%%%%%%%%%%%
\begin{figure}
\begin{center}
\epsfxsize=15cm \centerline{\epsffile{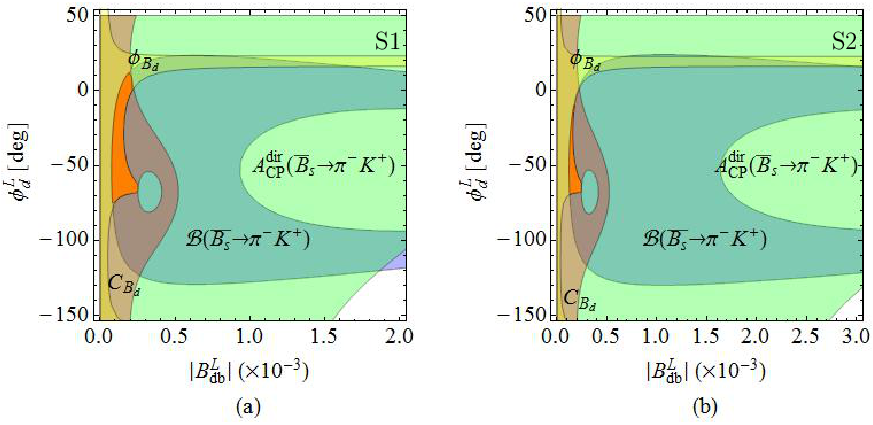}}
\centerline{\parbox{16cm}{\caption{\label{fitting}\small The allowed regions for the parameters $|B_{db}^L|$ and $\phi^L_d$  in S1~(a) and S2~(b) under the constraints from ${\cal B}(\bar{B}_s\to \pi^- K^+)$~(blue region), $A_{CP}^{dir}(\bar{B}_s\to \pi^- K^+)$~(green region), $C_{B_d}$~(brown region), $\phi_{B_d}$~(yellow region) and their combination (orange region) respectively. }}}
\end{center}
\end{figure}
%%%%%%%%%%%%%%%%%%%%%%%%%%%%%%%%%%%%%%%%%%%

With these theoretical formulae  and the  theoretical input parameters summarized in  Appendix, we now present our numerical analyses and discussions.
Our fitting is performed with that the experimental data is allowed within their $1.68\sigma$ ($\simeq\,90\%$ C.L.) error bars, while the theoretical uncertainties are also considered by varying the input parameters within their respective regions specified in the Appendix. In addition, we quote the scheme I to regulate the appearing end-point divergences. Since the $Z^{\prime}$ parameters $P_{ud}^{L,R}$ and $D_{ud}^{L,R}$ have been severely constrained by measured $B_s-\bar{B}_s$ mixing and $B\to\pi K$ decays~\cite{Chang2}, we demand that they cannot exceed their respective ranges listed in Table~\ref{inputsZp}, where S1 and S2 correspond to UTfit collaboration's two fitting results for $C_{B_s}$ and $\phi_{B_s}$~\cite{UTfit}.

With $|B_{db}^L|=0.15\times10^{-3}$ and the central values of theoretical input parameters, Fig.~\ref{fig3} shows the dependence of ${\cal B}(\bar{B}_s\to \pi^- K^+)$ on $\phi^L_d$, $P_{ud}^{L,R}$ and $D_{ud}^{L,R}$.
We find that ${\cal B}(\bar{B}_s\to \pi^- K^+)$ is sensitive to the $Z^{\prime}$ contributions induced by $P_{ud}^{R}$ and $D_{ud}^{L}$, but dull to $P_{ud}^{L}$ and $D_{ud}^{R}$. Interestingly, as Fig.~\ref{fig3}(b) and (c) shown, ${\cal B}(\bar{B}_s\to \pi^- K^+)$ could be reduced by $Z^{\prime}$ contributions with negative $P_{ud}^{R}$ and $D_{ud}^{L}$ at $\phi^L_d\sim-50^{\circ}$, which possibly presents a solution for the large ${\cal B}(\bar{B}_s\to \pi^- K^+)$ problem.

%%%%%%%%%%%%%%%%%%%%%%%%%%%%%%%%%%%%%%%%
\begin{table}
 \begin{center}
 \caption{The numerical results for the $Z^{\prime}$ couplings.}
 \label{fittingresult}
 \vspace{0.5cm}
 \small
 \doublerulesep 0.1pt \tabcolsep 0.05in
 \begin{tabular}{lccccccccccc} \hline \hline
 Solutions &$|B_{db}^L|(\times10^{-3})$ &$\phi^L_d(^{\circ})$ &$P_{ud}^{L}$   &$P_{ud}^{R}$   &$D_{ud}^{L}$      &$D_{ud}^{R}$\\\hline
 S1        &$0.16\pm0.08$               &$-33\pm45$           &$0.45\pm1.65$  &$-1.2\pm0.4$   &$-0.70\pm0.20$    &$-0.02\pm0.05$\\\hline
 S2        &$0.19\pm0.05$               &$-50\pm20$           &$-0.31\pm0.87$ &$-0.68\pm0.22$ &$-0.41\pm0.09$  &$-0.02\pm0.02$\\
 \hline\hline
 \end{tabular}
 \end{center}
 \end{table}
%%%%%%%%%%%%%%%%%%%%%%%%%%%%%%%%%%%%%%%%%%%%

With ${\cal B}(\bar{B}_s\to \pi^- K^+)$, $A_{CP}^{dir}(\bar{B}_s\to \pi^- K^+)$, $C_{B_d}$ and $\phi_{B_d}$ as constraints, corresponding to the two solutions for $P_{ud}^{L,R}$ and $D_{ud}^{L,R}$, the allowed regions for $|B_{db}^L|$ and $\phi_d^L$ are shown in Fig.~\ref{fitting} and the numerical results are listed in Table~\ref{fittingresult}. Interestingly, we find that the new  fitting result $\phi^L_d=-33^{\circ}\pm45^{\circ}$~($-50^{\circ}\pm20^{\circ}$) agrees well with the recent evaluation $\phi^L_d\sim-48^{\circ}$~\cite{Chang4} through the constraints from $B\to\pi\pi$.  As found in Refs.~\cite{Barger2,Chang2}, $B_d-\bar{B}_d$ mixing performs a very strong constraint on $B_{db}^L$, whose magnitude is suppressed to be $<0.24\times10^{-3}$ in both S1 and S2.
While, the lower bound $0.08(0.14)\times10^{-3}$ in S1(S2) for $|B_{db}^L|$ is dominated by ${\cal B}(\bar{B}_s\to \pi^- K^+)$. Unfortunately, the constraint from $A_{CP}^{dir}(\bar{B}_s\to \pi^- K^+)$ is
weak due to its large experimental error-bar. As shown in Table~\ref{inputsZp},  since  $P_{ud}^{L,R}$ in S2 are restricted to be smaller than their values in S1 case\cite{Chang2},  the lower bound for $|B_{db}^L|$ in S2 is demanded larger than the one in S1. As shown in Fig.~\ref{fitting}(a) and (b), the allowed ranges for both $|B_{db}^L|$ and $\phi_d^L$ in S2 are much smaller than the ones in S1.  So,  the solution S2 is much easier to be excluded by the future more precise measurements.  Due to the constraint from ${\cal B}(\bar{B}_s\to \pi^- K^+)$, the flavor-conserving $Z^{\prime}$ couplings $P_{ud}^{L,R}$ and $D_{ud}^{L,R}$ are further restricted. Their numerical results are also listed in Table~\ref{fittingresult}.

With the values of $Z^{\prime}$ couplings listed in Table~\ref{fittingresult} and the other inputs parameters given in the Appendix, including their uncertainties, we present our theoretical predictions for the observables in the fifth and sixth column of Table~\ref{br},~\ref{Acp} and \ref{mix}. We find ${\cal B}(\bar{B}_s\to \pi^- K^+)$ is reduced by $Z^{\prime}$ contributions. Due to large theoretical uncertainty,
its lower bound $5.8$($6.3$) $\times10^{-6}$ agrees with experimental measurement $(5.0\pm1.1)\times10^{-6}$ at $1\sigma$ error level in S1(S2).
As a  favored solution, within the allowed range for $Z^{\prime}$ couplings, we take
%%%%%%%%%%%%%%%%%%%%%%%%%%%%%%%%%%%%%%%%%%%%%%%%%
\begin{eqnarray}
|B_{db}^L|=0.24\times10^{-3}\,,\quad \phi^L_d=-55^{\circ}\,,\quad P_{ud}^{R}=-1.6\,(-0.9)\,,\quad D_{ud}^{L}=-0.90\,(-0.50)\,,
\end{eqnarray}
%%%%%%%%%%%%%%%%%%%%%%%%%%%%%%%%%%%%%%%%%%%%%%%%%
in S1(S2). With these  $Z^{\prime}$ values and the central values of the other parameters as inputs, we get
%%%%%%%%%%%%%%%%%%%%%%%%%%%%%%%%%%%%%%%%%%%%%%%%%
\begin{eqnarray}
{\cal B}(\bar{B}_s\to \pi^- K^+)=7.8\,(8.5)\times10^{-6}\,.
\end{eqnarray}
%%%%%%%%%%%%%%%%%%%%%%%%%%%%%%%%%%%%%%%%%%%%%%%%%
Comparing with the central value of the SM prediction $9.5\times10^{-6}$, we find ${\cal B}(\bar{B}_s\to \pi^- K^+)$ could be reduced by about $18\%$ ($11\%$) by $Z^{\prime}$ contributions at most. Such results means that S1 is more favored for the large  ${\cal B}^{th}(\bar{B}_s\to \pi^- K^+)$ problem than S2.
However, such result ${\cal B}(\bar{B}_s\to \pi^- K^+)=7.8\times10^{-6}$ in S1 is also $2\sigma$ larger than the experimental result $(5.0\pm1.1)\times10^{-6}$.

Our evaluations are based on the updated UTfit fitting results~\cite{UTfit} for $B_q-\bar{B}_q$ mixing.
Due to the constraint from $B_q-\bar{B}_q$ mixing, a CKM-like hierarchy is demanded in family non-universal $Z^{\prime}$ couplings, {\it i.e.}, $|B_{db}/B_{sb}|\sim|V_{td}^{\ast}V_{tb}/V_{ts}^{\ast}V_{tb}|\sim0.2$~\cite{Chang2}.
It is noted that both CDF and D0 collaborations have updated their measurements of  $B_s-\bar{B}_s$ mixing
recently based on $5.2fb^{-1}$ and $6.1fb^{-1}$ integrated luminosity, respectively. Their results read
%%%%%%%%%%%%%%%%%%%%%%%%%%%%%%%%%%%%%%%%%%%%%%%%%
\begin{eqnarray}
\label{CDFnewresult}
\beta_s&=&[0.02,0.52]\cup[1.08,1.55]\quad 68\% {\rm C.L.}\quad{\rm CDF}\,\cite{CDFnew}\,,\\
\label{D0newresult}
\phi_s^{J/\psi\phi}&=&-0.76^{+0.38}_{-0.36}\pm0.02\quad 68\% {\rm C.L.}\quad{\rm D0}\,\cite{D0new}\,.
\end{eqnarray}
%%%%%%%%%%%%%%%%%%%%%%%%%%%%%%%%%%%%%%%%%%%%%%%%%
For comparison, D0 result Eq.~(\ref{D0newresult}) could also be written as
%%%%%%%%%%%%%%%%%%%%%%%%%%%%%%%%%%%%%%%%%%%%%%%%%
\begin{eqnarray}
\label{D0newresult1}
\beta_s=0.38^{+0.18}_{-0.19}\pm0.01\quad 68\% {\rm C.L.}\quad{\rm D0}\,.
\end{eqnarray}
%%%%%%%%%%%%%%%%%%%%%%%%%%%%%%%%%%%%%%%%%%%%%%%%%
Unfortunately, these results have not been combined together yet by the two collaborations and by groups like
UTfit~\cite{UTfit}, CKMfitter~\cite{CKMfitter,ALenzaCKMfitter}, or HFAG~\cite{HFAG}.
It is interesting to note that, different from their former combined result $[0.27, 0.59]\cup[0.97, 1.30]$~\cite{CDFD0old}, SM expectation for $\beta_s$ $\sim0.018$ agrees with CDF measurement  Eq.~(\ref{CDFnewresult})
at $\sim1\sigma$ level.
D0 result in Eq.~(\ref{D0newresult1}), which is similar to their former result $[0.27, 0.59]\cup[0.97, 1.30]$~\cite{CDFD0old}, is also larger than SM prediction for $\beta_s$ at $\sim2\sigma$ level. So, obviously, CDF and D0 measurements have significantly difference with each other at small $\beta_s$ region, where is important for constraining  the strength of the NP contributions.

%%%%%%%%%%%%%%%%%%%%%%%%%%%%%%%%%%%%%%%%
\begin{figure}[t]
\begin{center}
\epsfxsize=7cm \centerline{\epsffile{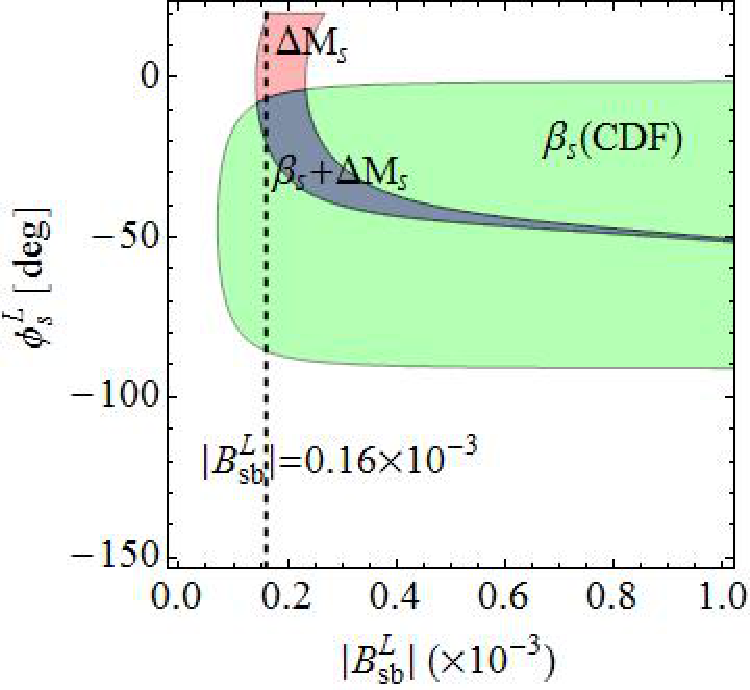}}
\centerline{\parbox{16cm}{\caption{\label{fig4}\small The allowed regions for the parameters $|B_{sb}^L|$ and $\phi^L_s$  under the constraints $\beta_s=[0.02,0.52](CDF)$~(green region), $\Delta M_s$~(pink region) and their combination~(blue region). The dashed line corresponds to the assumption $|B_{sb}^L|\approx|B_{db}^L|\approx0.16\times10^{-3}$ as a special scenario.}}}
\end{center}
\end{figure}
%%%%%%%%%%%%%%%%%%%%%%%%%%%%%%%%%%%%%%%%%%%

If $\beta_s\sim0.02$ is confirmed  by the coming experimental measurement, CKM-like hierarchy in $Z^{\prime}$ couplings matrix possibly could be violated, which implies that $|B_{sb}|$ could be as small as $|B_{db}|$. To confirm such a guess, we revisit $B_s-\bar{B}_s$ mixing within the family non-universal $Z^{\prime}$ model following the same track as Ref.~\cite{Chang2}. With $\beta_s=[0.02,0.52]$(CDF~Eq.~(\ref{CDFnewresult})) and  $\Delta M_s=17.77\pm0.12$~\cite{HFAG} as constraints and all of the central values of the parameters listed in appendix as input, the allowed regions for the new parameters $|B_{sb}^L|$ and $\phi^L_s$ are shown by Fig.~\ref{fig4}. So, from Fig.~\ref{fig4}, one may easily find that the relation $|B_{sb}|\sim|B_{db}|\sim0.2\times10^{-3}$, which is smaller than the former result $\sim1\times10^{-3}$~\cite{Chang2} by a factor about $1/5$, is allowed by the CDF updated measurement on $\beta_s$. Due to the constraints from $B\to\pi K$,~$\rho K$ and $\pi K^{\ast}$, the values of the combining parameters $\zeta^{LL,LR}_s$ and $\xi^{LL,LR}_s$, whose form could be derived through replacing $B_{db}^L$ in  Eq.~(\ref{comPara}) by $B_{sb}^L$, are unchangeable~\cite{Chang1,Chang2}. So, with $|B_{sb}|\sim|B_{db}|\sim0.2\times10^{-3}$ as input, after reevaluating the constraints from $B\to\pi K$ {\it et al.} (as Refs.~\cite{Chang1,Chang2} do), one may easily find that the results for $|P_{ud}^{L,R}|$ and $|D_{ud}^{L,R}|$ could be larger than the ones in Table~\ref{inputsZp} by a factor $~5$. So, as a possible special scenario, we take
%%%%%%%%%%%%%%%%%%%%%%%%%%%%%%%%%%%%%%%%%%%%%%%%%
\begin{eqnarray}
|B_{db}^L|=0.16\times10^{-3}\,,\quad \phi^L_d=-55^{\circ}\,,\quad P_{ud}^{R}=-1.2\times5\,,\quad D_{ud}^{L}=-0.70\times5\,,
\end{eqnarray}
%%%%%%%%%%%%%%%%%%%%%%%%%%%%%%%%%%%%%%%%%%%%%%%%%
for S1. With such $Z^{\prime}$ coupling values and the central values of other input parameters, we get ${\cal B}(\bar{B}_s\to \pi^- K^+)=5.1\times10^{-6}$ which is perfectly consistent with the experimental data $(5.0\pm1.1)\times10^{-6}$. Interestingly, as found in Ref.~\cite{Chang4}, $|B_{sb}|\sim|B_{db}|$ is also needed for resolving the ``$\pi\pi$ puzzle''. It also should be noted that our above analysis is based on the data $\beta_s\sim0.02$ is allowed, which is also needed to be confirmed by the future refined measurements of $B_s-\bar{B}_s$ mixing system at LHC-b and Tevatron.

\section{Conclusion}
In this paper, we have revisited
 $\bar{B}_s\to \pi K$, $\pi K^{\ast}$ and $\rho K$ decays within QCDF formalism.
We find that the large ${\cal B}^{th}(\bar{B}_s\to \pi^- K^+)$
problem could not be resolved by tuning the  annihilation contributions. In order to pursue possible solutions for the large  ${\cal B}^{th}(\bar{B}_s\to \pi^- K^+)$ problem, we evaluate the effects of a family non-universal $Z^{\prime}$ boson.  $B_d-\bar{B}_d$ mixing is
used to constrain the  $b-d-Z^{\prime}$ coupling. Our main conclusions are summarized as:

\begin{itemize}
\item The magnitude for $b-d-Z^{\prime}$ coupling $B_{db}^{L}$ is suppressed to $\sim0.2\times10^{-3}$ by
 $B_d-\bar{B}_d$ mixing. Based on the constrained $u-u-Z^{\prime}$ and $d-d-Z^{\prime}$ couplings by $B_s-\bar{B}_s$ mixing  and $B\to\pi K$ decays in Ref.~\cite{Chang2}, we find $P_{ud}^{L,R}$ and $D_{ud}^{L,R}$ are further restricted by ${\cal B}(\bar{B}_s\to \pi^- K^+)$, especially $P_{ud}^{R}$ and $D_{ud}^{L}$.

\item A new weak phase $\phi^L_d\sim-50^{\circ}$, negative $P_{ud}^{R}$ and/or $D_{ud}^{L}$ with larger absolute value are helpful to improve the agreement of ${\cal B}(\bar{B}_s\to \pi^- K^+)$ between the SM prediction and the experimental measurement. Compared the SM result, ${\cal B}(\bar{B}_s\to \pi^- K^+)$ could be reduced by about $18\%$($11\%$) at most in S1~(S2). S1 is more favored by the large ${\cal B}^{th}(\bar{B}_s\to \pi^- K^+)$ problem than S2.

\item The experimental measurement of  $\beta_s$ is very important to fix whether a CKM-like hierarchy is held in a family non-universal $Z^{\prime}$ couplings matrix. If the lower bound $0.02$ for $\beta_s$ measured by CDF is confirmed, $|B_{sb}|$ would  be as small as $|B_{db}|$. Within such a  scenario, ${\cal B}(\bar{B}_s\to \pi^- K^+)$ could be reduced to $5.1\times10^{-6}$ which agrees with the experimental data $(5.0\pm1.1)\times10^{-6}$.
\end{itemize}

The refined measurements for the $B_{s}$ non-leptonic
decay (especially $\bar{B}_s\to \pi^- K^+$) and $B_s-\bar{B}_s$ mixing in the LHC-b
 will provide a powerful  testing ground for the SM and possible NP scenarios. Our analysis
about the $Z^{\prime}$ effects on the observables for $\bar{B}_s\to \pi K$, $\pi K^{\ast}$, $\rho K$ decays and $B_q-\bar{B}_q$ mixing are useful for
probing or refuting the effects of a family non-universal $Z^{\prime}$ boson.

\section*{Acknowledgments}
The work is supported by the National Science Foundation under contract
No.11075059 and the Startup Foundation for Doctors of Henan Normal University under contract
No.1006.

\begin{appendix}
\section*{Appendix: Theoretical input parameters}

For the CKM matrix elements, we adopt the UTfit collaboration's
fitting results~\cite{UTfit,UTfitCKM}
\begin{eqnarray}
\overline{\rho}&=&0.132\pm0.02\,(0.135\pm0.04), \quad
\overline{\eta}=0.367\pm0.013\,(0.374\pm0.026),\nonumber\\
A&=&0.8095\pm0.0095\,(0.804\pm0.01),\quad
\lambda=0.22545\pm0.00065\,(0.22535\pm0.00065).
\end{eqnarray}
The values given in the bracket are the CKM parameters with assumption of the
 presence of generic New Physics,
which are used in our evaluations when the $Z^{\prime}$
contributions are included.

As for the quark masses, we take the current quark masses
\begin{eqnarray}
\frac{\overline{m}_s(\mu)}{\overline{m}_q(\mu)}&=&27.4\pm0.4\,~\cite{HPQCD:2006},\quad
\overline{m}_{s}(2\,{\rm GeV}) =87\pm6\,{\rm
MeV}\,~\cite{HPQCD:2006},
\quad\overline{m}_{c}(\overline{m}_{c})=1.27^{+0.07}_{-0.11}\,{\rm
GeV}~\cite{PDG08}\,\nonumber\\
\overline{m}_{b}(\overline{m}_{b})&=&4.20^{+0.17}_{-0.07}\,{\rm
GeV}~\cite{PDG08}\,,\quad
\overline{m}_{t}(\overline{m}_{t})=164.8\pm1.2\,{\rm
GeV}~\cite{PDG08}\,,
\end{eqnarray}
where $\overline{m}_q(\mu)=(\overline{m}_u+\overline{m}_d)(\mu)/2$,
and the difference between $u$ and $d$ quark is not distinguished.

The other one is the pole quark mass. In this paper, we
take~\cite{PDG08,PMass}
\begin{eqnarray}
 &&m_u=m_d=m_s=0, \quad m_c=1.61^{+0.08}_{-0.12}\,{\rm GeV},\nonumber\\
 &&m_b=4.79^{+0.19}_{-0.08}\,{\rm GeV}, \quad m_t=172.4\pm1.22\,{\rm GeV}.
\end{eqnarray}
and the decay constants
%%%%%%%%%%%%%%%%%%%%%%%%%%%%%%%%
\begin{eqnarray}
f_{B_{s}}&=&(231\pm15)~{\rm MeV}~~\cite{DecayCon}\,,f_{\pi}=(130.4\pm0.2)~{\rm MeV}~\cite{PDG08}\,,
f_{K}=(155.5\pm0.8)~{\rm MeV}~\cite{PDG08}\,,\nonumber\\
f_{K^{\ast}}&=&(217\pm5)~{\rm MeV}~\cite{BallZwicky}\,,
f_{K^{\ast}}^T(2.2GeV)=(156\pm10)~{\rm MeV}~\cite{BallZwicky}\,,\nonumber\\
f_{\rho}&=&(209\pm2)~{\rm MeV}~\cite{BallZwicky}\,,\nonumber
f_{\rho}^T(2.2GeV)=(147\pm10)~{\rm MeV}~\cite{BallZwicky}.
\end{eqnarray}
%%%%%%%%%%%%%%%%%%%%%%%%%%%%%%%%
As for the B-meson lifetimes,
$\tau_{B_{s}}=1.470\,{\rm ps}~\cite{PDG08}$
is used. We take the heavy-to-light transition form
factors
%%%%%%%%%%%%%%%%%%%%%%%%%%%%%%%%%
\begin{eqnarray}
 A^{\bar{B}_s\to {K^{\ast}}}_{0}(0)=0.360\pm0.034~\cite{BallZwicky}\,,
 F^{\bar{B}_s\to {K}}_{0}(0)=0.30^{+0.04}_{-0.03}~\cite{Duplancic}\,.
\end{eqnarray}
%%%%%%%%%%%%%%%%%%%%%%%%%%%%%%%%
\end{appendix}

 \end{document}